\documentclass{ws-ijmpa-hepth0403037}   

\newcommand {\CSlike}{Chern--Simons-like}
\newcommand {\CS}    {Chern--Simons}
\newcommand {\PV}    {Pauli--Villars}
\newcommand {\LC}    {Levi--Civita}
\newcommand {\BN}    {Bogoliubov--Nambu}
\newcommand {\WZNW}  {Wess--Zumino--Novikov--Witten}
\newcommand {\BCS}   {Bardeen--Cooper--Schrieffer}
\newcommand {\cgth}  {chiral gauge theory}
\newcommand {\qfth}  {quantum field theory}
\newcommand {\qfths} {quantum field theories}
\newcommand {\rqfth} {relativistic quantum field theory}

\def\identity{{\rm 1\kern-.12em \rule{0.3pt}{1.5ex}\raisebox{0.0ex}
                  {\rule{0.1em}{0.3pt}}}}

\begin{document}

\markboth{F. R. Klinkhamer and G. E. Volovik}
{Emergent CPT violation from the splitting of Fermi points}

%
\catchline{}{}{}{}{}
%

\title{Emergent CPT violation from the splitting of Fermi points}

\author{\footnotesize F. R. KLINKHAMER}

\address{Institute for Theoretical Physics, University of Karlsruhe (TH),
         76128 Karlsruhe, Germany\\frans.klinkhamer@physik.uni-karlsruhe.de}

\author{G. E. VOLOVIK}

\address{Helsinki University of Technology,
Low Temperature Laboratory,\\ P.O. Box 2200, FIN--02015 HUT,  Finland, \\
Landau Institute for Theoretical Physics, 119334 Moscow, Russia\\
volovik@boojum.hut.fi}

\maketitle

\pub{Received 6 September 2004}{}

\begin{abstract}
In a fermionic quantum vacuum, the parameters $k_\mu$
of a CPT-violating Chern--Simons-like action term
induced by CPT-violating parameters
of the fermionic sector depend on the universality class of the system.
As a concrete example, we consider the Dirac Hamiltonian of a
massive fermionic quasiparticle and add a
particular term with purely-spacelike CPT-violating parameters
$b_\mu=(0,{\bf b})$.
A quantum phase transition separates two phases, one with a
fully-gapped fermion spectrum  and the other with
topologically-protected Fermi points (gap nodes).
The emergent Chern--Simons ``vector''
$k_\mu=(0,{\bf k})$ now consists of two parts. The regular part,
${\bf k}^{\rm reg}$,
is an analytic function of $|{\bf b}|$
across the quantum phase transition and may be nonzero due to explicit
CPT violation at the fundamental level.
The anomalous (nonanalytic) part,
${\bf k}^{\rm anom}$,
comes solely from the Fermi points and is proportional to their
splitting.
In the context of condensed-matter physics, the quantum phase transition
may occur in the region of the BEC--BCS crossover for Cooper pairing in the
$p$--wave channel.
For elementary particle physics,
the splitting of Fermi points may lead to neutrino oscillations,
even if the total electromagnetic Chern--Simons-like term cancels out.

\end{abstract}

\section{Introduction}
\label{sec:Introduction}

Condensed-matter physics provides us with a broad class of
\qfths~which are not restricted by Lorentz invariance. This
allows us to consider many problems in the \rqfth~of the
electroweak Standard Model\cite{EWSM} from a more general perspective.
Indeed, general consideration of \qfths~have revealed that their
basic properties are determined by momentum-space topology, which classifies
the vacua of \qfth~according to three universality classes.

In the present article, we apply the general topological argument to
the particular case of a CPT-violating Chern--Simons-like term.
For this case, certain calculations in the framework of
\rqfth~have turned out to be
problematic because of an ambiguity tracing back to
different regularization schemes.
We find, on the other hand, that the Chern--Simons parameter can be expressed
purely  by a topological invariant in momentum space. In fact, this
approach provides a topological regularization scheme which is more
general than other schemes based on symmetry.

The topological approach holds for two common applications of
\rqfth, namely as a description of fundamental physics
or as an emergent phenomenon of a fermionic quantum vacuum
with the known elementary particles corresponding
to the quasiparticles of the system.
Since the topological approach has been proven correct in the class of
condensed-matter systems with emerging \rqfth~in the low-energy corner,
we expect that it may be relevant to Standard Model physics as well.

The momentum-space topology of the fermionic spectrum determines
the universality classes of all known quantum vacua in which momentum
is a well defined quantity (i.e., which have translation invariance);
cf. Chapters 7 and 8 of Ref.~\refcite{VolovikBook}.
But the vacu\-um of the Stan\-dard Mo\-del above the electroweak transition
(with vanishing fermion masses)
is marginal and, strictly speaking, belongs to all
three universality classes. The Standard Model vacuum above the
electroweak transition has, in fact, a
multiply degenerate Fermi point ${\bf p}=0$ and the total topological
charge of this point is zero. The implication is that
momentum-space topology cannot protect this vacuum against
decay into other (topologically-stable) vacua.

The Standard Model vacuum above the electroweak transition is only
protected by symmetries, namely the continuous electroweak symmetry, the
discrete symmetry discussed in Section~12.3.2
of Ref.~\refcite{VolovikBook}, and the CPT symmetry.
Explicit violation or spontaneous breaking of these symmetries
(CPT in particular) leads to
three  possible scenarios, corresponding to the three universality
classes of fermionic vacua:
\par
(i) \emph{Annihilation of Fermi points---} Fermi points with
opposite topological charge merge and disappear.
[Fermi points (gap nodes)
${\bf p}_a$ are points in three-dimensional momentum space at
which the energy spectrum $E({\bf p})$ of the fermionic quasiparticle
has a zero, i.e., $E({\bf p}_a)=0$. Some early references on Fermi points
include
Refs.~\refcite{AbrikosovBeneslavskii,LeggettTakagi,NielsenNinomiya}
and \refcite{VolovikMineev}.]
The annihilation of Fermi points allows the
fermions to  become massive, as happens with
the quarks and charged leptons of the Standard Model below the
electroweak transition.
\par
(ii) \emph{Splitting of Fermi points---} Fermi
points split and separate along a spatial direction in
four-dimensional energy-momentum space.  The topological charges of
the new isolated Fermi points are nonzero and the corresponding vacua
are topologically protected. For Standard Model physics, the
splitting of Fermi points  gives rise to a CPT-violating Abelian
\CSlike~term\cite{CarollFieldJackiw,AdamKlinkhamerNPB}
\begin{equation}
S_{\rm \,CS-like}= \int d^4x \; k_\mu\,\epsilon^{\mu\nu\rho\sigma}
A_\nu(x)\:\partial_\rho  A_\sigma(x)~,
\label{CSliketerm}
\end{equation}
with gauge field
$A_\mu(x)$, \LC~symbol $\epsilon^{\mu\nu\rho\sigma}$, and
a purely spacelike ``vector'' $k_\mu=(0,{\bf k})$.
As will be explained later, the
Fermi-point origin of such a \CSlike~term is, in a way,
complementary to the mechanism of the CPT
anomaly\cite{CPTanomaly,FundamentalTimeAsymmetry,RigorousResult,SpacetimeFoam}
in \rqfth. For condensed-matter physics, the prime example is
superfluid $^3$He--A; cf. Refs.~\refcite{VollhardtWoelfle}
and \refcite{ExoticProperties}.
Here, the splitting is extremely large, with Fermi points
separated by a ``Planck-scale'' momentum.
But the splitting can, in principle, be controlled by a
tunable interaction in the $p$--wave Cooper channel (see below).
\par
(iii)
\emph{Formation of Fermi surfaces---} Fermi points split
along the energy axis in four-dimensional energy-momentum space and give
rise to a Fermi surface in three-dimensional momentum space.
[Fermi surfaces $S_a$ are two-dimensional surfaces in
three-dimensional momentum space on which the energy spectrum $E({\bf
p})$ is zero, i.e., $E({\bf p})=0$ for all ${\bf p}  \in S_a$.]
For Standard Model physics, this scenario would lead to a
\CSlike~term (\ref{CSliketerm}) with a purely timelike
``vector'' $k_\mu=(k_0,{\bf 0})$.
For condensed-matter physics, this scenario need not be
prohibited by the helical instability of the
vacuum, since the total $k_0$ obtained by summation over all Fermi
surfaces can be zero due to the remaining symmetries of the
fermions.\cite{VolovikCPT}

The choice between these three scenarios depends on which
symmetry-breaking
mechanism is stronger. The important point is that there are only three
possibilities corresponding to the three universality classes of
fermionic vacua:
(i) vacua with
fully-gapped fermionic excitations;
(ii) vacua with fermionic
excitations characterized by Fermi points (these excitations behave
as Weyl fermions close to the Fermi points); (iii)  vacua
with fermionic excitations characterized by Fermi surfaces.

Experimentally, the quantum phase transition between vacua of different
universality classes can be investigated with laser-manipulated fermionic
gases. The idea is to probe the crossover between the weak-coupling state
described by \BCS~(BCS) theory and the
strong-coupling limit  described by Bose--Einstein condensation (BEC) of
fermionic atom pairs. If the pairing occurs in the $p$--wave state,
the BEC--BCS
crossover may be accompanied by a quantum phase transition between
vacua with Fermi points and fully-gapped vacua.
Starting from the BEC phase,
a marginal Fermi point is formed at the quantum phase transition,
which splits into
topologically-stable Fermi points as the interaction strength is
reduced further (BCS phase).

The rest of the article elaborates these points.
Section~\ref{sec:quantumphasetransition} explains the possible origin of
CPT violation as a quantum phase transition which splits a
multiply degenerate Fermi point and
Section~\ref{sec:CSliketerm} discusses the induced \CSlike~term
with a purely spacelike vector.
(\ref{sec:Appendix} gives the \CS~vector in the form of a
momentum-space topological invariant.)
Section~\ref{sec:WZNW3HeA} describes this action term in the context of
superfluid $^3$He--A. Section~\ref{sec:p-wavesuperconductors}
considers the role of Fermi points for $p$--wave
superconductors and the Standard Model (the possibility of a new
mechanism for neutrino oscillations is also pointed out).
Section~\ref{sec:Timelikeparameter} discusses the
origin of a different type of \CSlike~term with a purely timelike
vector, as being due to the appearance of a Fermi surface
(another possible consequence being again neutrino oscillations).
Section~\ref{sec:finite-size} mentions two other mechanisms for the
splitting of Fermi points (or appearance of Fermi surfaces),
namely finite-size effects and the presence of defects.
Section~\ref{sec:Conclusion} presents some concluding remarks
and briefly discusses  experiments of the BEC--BCS crossover.

\section{CPT Violation as Quantum Phase Transition}
\label{sec:quantumphasetransition}

As a typical example of the spacelike splitting of multiply degenerate
Fermi points, we consider the marginal Fermi point of the Standard Model
above the electroweak transition.
In this section and the next, natural units are employed with $c=\hbar=1$.

Take, for simplicity, a single pair of relativistic chiral
fermions, that is, one right-handed fermion and one left-handed fermion.
These are Weyl fermions with Hamiltonians
$H_{\rm right}=\vec{\sigma}\cdot{\bf p}$ and
$H_{\rm left}=-\vec{\sigma}\cdot{\bf p}$,
where $\vec{\sigma}$ denotes the triplet of Pauli matrices.
Each of these Hamiltonians has a topologically-stable Fermi
point ${\bf p}=0$, where the Hamiltonian is zero.

For a general system, relativistic or
nonrelativistic, the stability of the
$a$-th Fermi point is
guaranteed by the topological invariant $N_{a}$,
which can be written
as a surface integral in energy-momentum space. In terms
of the fermionic propagator $G(ip_0,{\bf p})$,
the topological invariant is\cite{VolovikBook}
\begin{equation}
N_{a} = {1\over{24\pi^2}}\,\epsilon_{\mu\nu\rho\sigma}~
{\rm tr}\, \oint_{\Sigma_a} dS^{\sigma} \;
G\frac{\partial}{\partial p_\mu} G^{-1}\;G
\frac{\partial}{\partial p_\nu}  G^{-1}\;G  \frac{\partial}{\partial
p_\rho}G^{-1}~,
\label{TopInvariant}
\end{equation}
where $\Sigma_a$
is a three-dimensional surface around the isolated Fermi point
$p_{\mu a}=(0,{\bf p}_a)$ in four-dimensional energy-momentum
space and `tr' stands for the trace over the spin indices.

In our case, we have two Weyl fermions, $a=1$ for the right-handed fermion
and $a=2$ for the left-handed one. The corresponding Green's functions are
given by
\begin{equation}
G^{-1}_{\rm right}(ip_0,{\bf p})=ip_0
-\vec{\sigma}\cdot{\bf p}\;,\quad
G^{-1}_{\rm left} (ip_0,{\bf
p})=ip_0 +\vec{\sigma}\cdot{\bf
p}~.
\label{GreenFWeyl}
\end{equation}
The positions of the Fermi points coincide, ${\bf p}_1={\bf p}_2=0$,
but their topological charges (\ref{TopInvariant}) are different,
$N_1=+1$ and $N_2=-1$.
For this simple case, the topological charge equals the
chirality of the fermions, $N_a=C_a$.

The splitting of coinciding Fermi points can be described by
the Hamiltonians
$H_{\rm right}=\vec{\sigma}\cdot({\bf
p}-{\bf p}_1)$ and
$H_{\rm left}=-\vec{\sigma}\cdot({\bf p}-{\bf
p}_2)$, with
${\bf p}_1=-{\bf p}_2 \equiv {\bf b}$ from momentum conservation.
The  real vector ${\bf b}$ is assumed to be odd under CPT,
which introduces CPT violation into these Hamiltonians.
The $4\times 4$ matrix of the combined Green's function has the form
\begin{equation}
G^{-1}(ip_0,{\bf p}) = \left(\matrix{ip_0 -
\vec{\sigma}\cdot({\bf
p}-{\bf b})&0\cr
0&ip_0+
\vec{\sigma}\cdot({\bf p}+{\bf b})\cr } \right) ~.
\label{ModifiedGreenWeyl}
\end{equation}
   From Eq.~(\ref{TopInvariant}) follows that ${\bf p}_1={\bf b}$ is
the Fermi point with topological charge $N_1=+1$ and
${\bf p}_2=-{\bf b}$
the Fermi point  with topological charge $N_2=-1$.

\begin{figure}
\centerline{\psfig{file=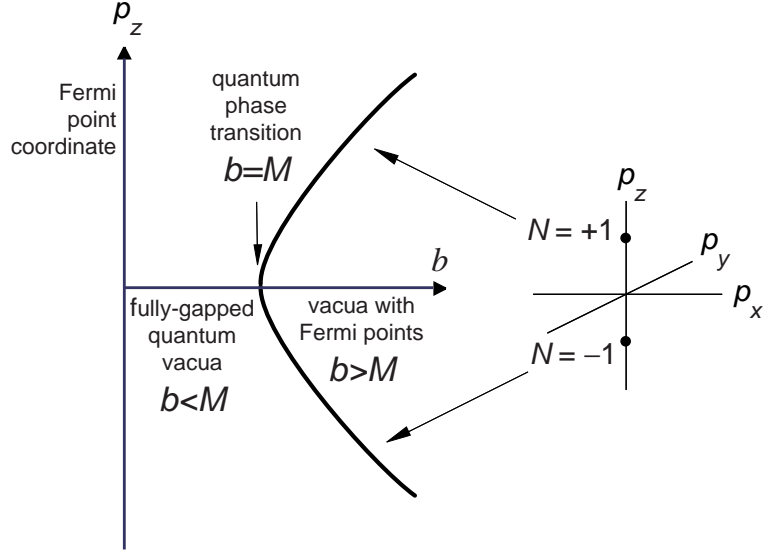,width=10cm}} 
\vspace*{8pt}
\caption{Quantum phase transition at $b \equiv |{\bf b}|=M$ between
anomaly-free, fully-gapped vacua for $b<M$ and vacua with
topologically-protected Fermi points for $b>M$, where ${\bf b}$ is a
CPT-odd parameter of the  modified Dirac Hamiltonian (\ref{ModifiedHDirac}).
The two Fermi points are characterized by
topological invariants $N = \pm 1$, as indicated by the inset on the right
(the unit vector $\widehat{\bf z}$ in momentum space is parallel to ${\bf b}$).
These Fermi points give an anomalous (nonanalytic)
contribution to the induced Chern-Simons parameter ${\bf k}$.}
\label{Fig1}
\end{figure}

Physically, it is perhaps more instructive to discuss the appearance and
splitting of Fermi points by starting from a Dirac fermion
(fermionic quasiparticle) with rest energy  $M$.
Consider, then, a  single Dirac fermion in the presence of a CPT-violating
vector ${\bf b}$, with a modified Hamiltonian ($c=1$)
\begin{equation}
H=\left(\matrix{\vec{\sigma}\cdot({\bf p}-{\bf b})&M\cr
      M&-
\vec{\sigma}\cdot({\bf p}+{\bf b})\cr } \right)
=H_{\rm Dirac} -
\identity_{\,2}  \otimes
  \left(\vec{\sigma}\cdot{\bf b}\right) ~.
\label{ModifiedHDirac}
\end{equation}
This is the typical starting point for investigations of the effects of
CPT violation in the fermionic sector;
see, e.g., Refs.~\refcite{ColladayKostelecky}, \refcite{PerezJHEP}, and
\refcite{Lehnert}.
The energy spectrum of the Hamiltonian (\ref{ModifiedHDirac}) is
\begin{equation}
E^2_\pm= M^2+|{\bf
p}|^2+b^2
            \pm 2\,\sqrt{b^2\,M^2+({\bf p}\cdot{\bf b})^2}
~,
\label{ModifiedEnergyDirac}
\end{equation}
with $b\equiv |{\bf b}|$.

Allowing for variable $b$ in Eqs.~(\ref{ModifiedHDirac})
and (\ref{ModifiedEnergyDirac}),
one finds a quantum phase transition at $b=M$ between
fully-gapped vacua for $b<M$ and vacua with two Fermi points for $b>M$.
These Fermi points are given by
\begin{equation}
{\bf p}_1=  {\bf b}\,\sqrt{1-{M^2 / b^2}}  \;,\qquad
{\bf p}_2=- {\bf b}\,\sqrt{1-{M^2 / b^2}}   ~.
\label{FPDiracFermions}
\end{equation}
       From Eq.~(\ref{TopInvariant}),
now with a trace over the indices of the $4\times 4$ Dirac matrices,
follows that ${\bf p}_1$ is
the Fermi point with topological charge $N_1=+1$ and
${\bf p}_2$ the Fermi point with topological charge $N_2=-1$
(see Fig.~\ref{Fig1}).  The magnitude of the splitting of the
two Fermi points is $2\,\sqrt{b^2-M^2}\,$. At the quantum phase
transition between the vacua of the two universality classes  (i.e., at
$b=M$), the Fermi points with
opposite charges annihilate each other and form a marginal Fermi point
${\bf p}_0 =0$. The momentum-space topology of this marginal
Fermi point is trivial (topological invariant $N_0=+1-1=0$).

\section{Emergent Chern--Simons-like Term}
\label{sec:CSliketerm}

Let us now consider how the splitting of Fermi points in the fermionic
sector induces an anomalous action term in the gauge-field sector.
Start from the spectrum of a single electrically charged Dirac fermion
(charge $q$) and again set $c=\hbar=1$.
In the presence of the vector potential ${\bf A}$ of a $U(1)$
gauge field, the minimally-coupled Hamiltonian is
\begin{equation}
H=\left(\matrix{
       \vec{\sigma}\cdot({\bf p}-q{\bf A} -{\bf b})&M\cr
M&- \vec{\sigma}\cdot({\bf p}-q{\bf A}+{\bf b})\cr }\right)   ~.
\label{ModifiedHDiracGauge}
\end{equation}
The positions of the Fermi points for $b \equiv |{\bf b}| > M$ are
shifted due to the gauge field,
\begin{equation}
{\bf p}_a=q{\bf A}\pm {\bf b}\;\sqrt{1-M^2 /b^2}    ~,
\label{FPWithGauge}
\end{equation}
with a plus sign for $a=1$ and a minus sign for $a=2$.
This result  follows immediately from
Eq.~(\ref{FPDiracFermions}) by the minimal substitution
${\bf p}_a \rightarrow {\bf p}_a - q {\bf A}$,
consistent with the gauge principle.

At this moment, we can simply use the general expression for the
Wess--Zumino--Novikov--Witten  (WZNW) action induced by Fermi points,
which holds for any system.
The WZNW action is represented by the following sum over Fermi points
[see, e.g., Eq.~(6a) in Ref.~\refcite{VolovikKonyshev1988}]:
\begin{equation}
S_{\rm WZNW}=-{1\over 12\pi^2}\, \sum_a\,N_{a}\int d^3x~d t~d\tau \;
{\bf p}_a\cdot(\partial_t{\bf p}_a\times\partial_\tau{\bf p}_a)\;,
\label{WessZuminoGeneral}
\end{equation}
where $\tau \in [0,1]$ is an additional coordinate which
parametrizes a disc, with the usual spacetime at the boundary $\tau=1$.
The normalization factor $1/12\pi^2$ of this anomalous action
is determined by a topological argument
similar to that in Ref.~\refcite{Witten1983}; see Eqs.~(3.10)
and (3.11) of Ref.~\refcite{Volovik1993}.

For \rqfth~and with different charges $q_a$ at
the different Fermi points,  the dependence on ${\bf A}$ of the
expression (\ref{WessZuminoGeneral})
comes  from the shifts of the Fermi points,
${\bf p}_a={\bf p}_a^{(0)}+  q_a(\tau)\, {\bf A}$. Here, a parametrization
is used in which the charges $q_a(\tau)$ are zero at the center of
the disc, $q_a(0)=0$, and equal to the physical charges at the boundary of
the disc, $q_a(1)=q_a$.  From Eq.~(\ref{WessZuminoGeneral}),
one obtains the general form for the Abelian \CSlike~term
\begin{equation}
S_{\rm CS-like}={1\over 24\pi^2}\,\sum_a N_{a}\,q_a^2\,
\int d^3x\, d t \;  {\bf p}_a^{(0)} \cdot ({\bf A}\times
\partial_t{\bf A}) ~.
\label{CSgeneral}
\end{equation}
This result has the ``relativistic'' form (\ref{CSliketerm})
with a purely spacelike ``vector''
\begin{equation}
k_\mu=(0,{\bf k})=
\left(0,\, (24\pi^2)^{-1} \sum_a\,{\bf p}^{(0)}_a\,q_a^2\,N_a \right)~.
\label{ManyFermiPointsSpaceLike}
\end{equation}
Note that only gauge invariance has been assumed in the derivation
of Eq.~(\ref{ManyFermiPointsSpaceLike}).
As shown in \ref{sec:Appendix}, the \CS~vector
(\ref{ManyFermiPointsSpaceLike}) can be written in the form
of a momentum-space topological invariant.

Returning to the case of a single Dirac fermion with charge $q$ and
using Eqs.\ (\ref{ManyFermiPointsSpaceLike}) and (\ref{FPDiracFermions}),
one finds that the CPT-violating \CS~parameter
${\bf k}$ can be expressed in terms of the  CPT-violating
parameter ${\bf b}$ of the fermionic sector,
\begin{equation}
{\bf k}={q^2\over 12\pi^2}\;\theta(b - M)\;{\bf b}\;\sqrt{1-M^2/ b^2}  ~.
\label{SpaceLikeKAphase}
\end{equation}
This contribution to ${\bf k}$ comes from the splitting of a marginal
Fermi point and requires $b > M$, as
indicated by the step function on the right-hand side
[$\,\theta(x)=0$ for $x \leq 0$  and $\theta(x)=1$ for $x>0\,$].

In the context  of \rqfth,
the existence of such a nonanalytic contribution to ${\bf k}$
has also been found by Perez-Victoria\cite{PerezPRL} and Andrianov
\emph{et al.}\cite{AndrianovJHEP} using
standard regularization methods, but with a prefactor larger by
a factor $3$ and $3/2$, respectively.
The result (\ref{SpaceLikeKAphase}), on the other hand, is determined by
the general topological properties of the Fermi points and applies
to  nonrelativistic \qfth~as well. In condensed-matter \qfth, the
result was obtained without ambiguity, since the microphysics is
known at all scales and the regularization occurs naturally.

The induced \CS~vector (\ref{SpaceLikeKAphase}) from explicit CPT violation
in the Dirac Hamiltonian (\ref{ModifiedHDiracGauge}) differs from the
anomalous result in chiral gauge theory over topologically
nontrivial spacetime
manifolds.\cite{CPTanomaly,FundamentalTimeAsymmetry,RigorousResult,SpacetimeFoam}
Most importantly, the result (\ref{SpaceLikeKAphase}) holds
for a vectorlike gauge theory (with a Dirac fermion), whereas
the CPT anomaly appears exclusively in chiral gauge theories (with
Weyl fermions). There are, however, also similarities. First,
the result (\ref{SpaceLikeKAphase}) arises only from a
strong enough source of CPT violation ($b \equiv |{\bf b}| > M$)
but can have an arbitrarily small mass scale
($2\,\sqrt{b^2-M^2}\,$). This behavior resembles that of
the CPT anomaly in the usual form, which requires a large enough
ultraviolet cutoff (\PV~mass or inverse lattice spacing) but
can have an arbitrarily small mass scale
($1/L$, for size $L$ of the compact dimension with periodic spin structure).
Second, the pattern of two separated Fermi points is reminiscent of
the different options for the definition of the fermion measure
in the relevant chiral lattice gauge theory, where the choice of one or the
other option leads to the CPT anomaly
(see, in particular, Section~5.3 of Ref.~\refcite{RigorousResult}).

\section{WZNW Action for Superfluid $^3$He--A}
\label{sec:WZNW3HeA}

The \BN~Hamiltonian which qualitatively describes fermionic
quasiparticles in $^3$He--A is given by
\begin{eqnarray}
H=\left(\matrix{ p^2/ 2m_3  -\mu
&c_\perp\,{\bf p}\cdot (\widehat{\bf e}_1+ i\, \widehat{\bf e}_2) \cr
   c_\perp\,{\bf p}\cdot (\widehat{\bf e}_1- i\, \widehat{\bf e}_2)
&-p^2/ 2m_3 +\mu \cr }
\right) \;,
\label{BogoliubovNambuH}
\end{eqnarray}
with the mass $m_3$ of the $^3$He atom, the orthonormal triad
$(\widehat{\bf e}_1,\, \widehat{\bf e}_2,\,
\widehat{\bf l}=\widehat{\bf e}_1\times \widehat{\bf e}_2)$, and
the maximum transverse speed $c_\perp$ of the quasiparticles.
The unit vector $\widehat{\bf l}$ corresponds to
the direction of the orbital momentum of the Cooper pairs.
The energy spectrum of these  \BN~fermions is
\begin{equation}
E^2= \left({p^2\over 2m_3}-\mu\right)^2+
         c_\perp^2\,\left({\bf p}\times \widehat{\bf l}\,\right)^2  ~.
\label{BogoliubovNambuE}
\end{equation}

This last expression for the energy makes clear that
Fermi points (with $E=0$) only
occur for the case of positive chemical potential, $\mu>0$.
These Fermi points are ${\bf p}_1= p_F \, \widehat{\bf l}$ and
${\bf p}_2=-p_F \, \widehat{\bf l}$, with Fermi momentum
$p_F=\sqrt{2\mu\, m_3}$.
     From the general expression (\ref{WessZuminoGeneral}),
one finds the following \WZNW~(WZNW) action for superfluid $^3$He--A
(cf. Refs.~\refcite{ExoticProperties} and \refcite{VolovikKonyshev1988}):
\begin{equation}
S^{\rm anomalous}_{\rm WZNW}=
-{\hbar K_0\over 2}\int d^3x~d t~d\tau \; \widehat{\bf l}\cdot
\left(\,\partial_t\widehat{\bf l}\times\partial_\tau\widehat{\bf l}\,\right)~,
\label{WessZuminoHe}
\end{equation}
with  coefficient $K_0 \equiv (p_F/\hbar)^3/3\pi^2$.
This term has a nonanalytic dependence on $\mu$ through $p_F$
and plays a crucial role in the orbital dynamics of $^3$He--A.

The effective gauge field, which emerges for ``relativistic''
fermions  in the vicinity of the Fermi points, is the collective mode of
the shift of the Fermi points, ${\bf A}'=p_F\,\widehat{\delta {\bf l}}$, where
$\widehat{\delta {\bf l}}$ is a perturbation around the vacuum value,
$\widehat{\bf l}=\widehat{\bf l}_0 + \widehat{\delta {\bf l}}$.
Expanding Eq.~(\ref{WessZuminoHe}) in terms of
${\bf A}' \equiv \hbar\,{\bf A}  = p_F\,\widehat{\delta {\bf l}}$,
one finds that the effective \CSlike~term in $^3$He--A has the
``relativistic'' form (\ref{CSliketerm}) with spacelike ``vector''
\begin{equation}
k_\mu=(0, {\bf k})=
\left(\,0,\,(12\pi^2)^{-1}\,p_F\, \widehat{\bf l}_0\,\right)\;.
\label{kmu3HeA}
\end{equation}
   As mentioned above, the vector $\widehat{\bf l}_0$
gives the direction of the orbital momentum of the Cooper pairs.
The nonvanishing angular momentum (a T-odd vector)
makes clear that time-reversal symmetry is spontaneously broken
in the quantum vacuum of $^3$He--A . (For relativistic quantum
field theory which emerges in the vicinity of Fermi points, this
implies T and CPT violation for the quasiparticles, e.g., the photon and
the electron. Indeed, it is possible to construct a light clock which
ticks differently if the velocities are
reversed.\cite{FundamentalTimeAsymmetry})

For $^3$He--A, there is also a regular
part\cite{ExoticProperties,VolovikKonyshev1988} of the WZNW action,
\begin{equation}
S^{\rm regular}_{\rm WZNW}=
\int d^3x~d t~d\tau \;{\hbar n\over 2}\;\widehat{\bf l}\cdot
\left(\,\partial_t\widehat{\bf l}\times\partial_\tau\widehat{\bf l}\,\right)~,
\label{WessZuminoHeReg}
\end{equation}
with $n$ the particle density of $^3$He atoms.
This regular part comes from the angular momentum of the liquid
(angular momentum density ${\bf L}=\hbar\,\widehat{\bf l}\;n/2\,$),
whereas the nonanalytic contribution (\ref{WessZuminoHe})
comes from the Fermi points and is proportional to their
splitting ($2p_F \, \widehat{\bf l}\,$).
(In bulk nonsuperconducting materials, there may be
a similar nonanalytic contribution to the Hall conductivity;
see \ref{sec:Appendix}.)
The regular contribution does not depend on the
presence of Fermi points and remains
when the Fermi points disappear by a quantum phase transition.

The disappearance of Fermi points
takes place for strong interactions in the Cooper channel,
when the quantum phase transition to the fully-gapped state occurs.
Starting from the Fermi-point phase ($\mu > 0$) and increasing
the interaction strength, the value of $\mu$ drops and, at a
particular  value of the interaction strength,
the momentum-space topology changes.  According to
Eq.~(\ref{BogoliubovNambuE}),
the two Fermi points annihilate each other at
$\mu=0$ and disappear for $\mu<0$; cf. Sections~6.2 and 6.5 of
Ref.~\refcite{ExoticProperties}. The quantum phase transition at
$\mu=0$  may be called a Lifshitz transition, by analogy with
the quantum phase transitions in metals at which the momentum-space topology
of the Fermi surface changes. Fermi points give rise to the
chiral anomaly and to anomalous (nonanalytic) terms in the orbital
dynamics. These anomalies are absent in the fully-gapped state ($\mu <0$).

Real $^3$He--A lives, however, in the  $\mu>0$ region of the phase
diagram.   A fully-gapped vacuum on the other side of the phase boundary
(that is, a vacuum without
Fermi points, but with the same breaking of time-reversal symmetry) can
perhaps be obtained in laser-manipulated Fermi gases; see
Section~\ref{sec:Conclusion}  for further remarks.

\section{Fermi Points for $p$--Wave Superconductors and Standard Model}
\label{sec:p-wavesuperconductors}

The vector ${\bf k}$ of the Abelian
\CSlike~term induced by all Fermi points is
given by Eq.~(\ref{ManyFermiPointsSpaceLike}). Considering massless
fermions with charges $q_a$ and taking into account that the
topological charge $N_a$ for Weyl fermions coincides with their chirality
$C_a=\pm 1$ (see Section~\ref{sec:quantumphasetransition}), one obtains
\begin{equation}
{\bf k}={1\over 24\pi^2}\, \sum_a\,{\bf p}_a\, q_a^2\,C_a~,
\label{ManyFermiPointsSpaceLike2}
\end{equation}
where the integer $a$ labels the Fermi points ${\bf p}_a$.

For $^3$He--A, the sum of Eq.~(\ref{ManyFermiPointsSpaceLike2}) gives
$\sum_a\,{\bf p}_a\,q_a^2\,C_a=2p_F\,\widehat{\bf l}_0$
and the induced vector ${\bf k}$ is nonzero.
But, for other vacua, the induced vector ${\bf k}$ can be zero after
summation.  An example of this cancellation
is provided by the so-called $\alpha$--phase of spin-triplet pairing
in superconductors; cf. Ref.~\refcite{VollhardtWoelfle}.
This phase contains eight Fermi points ${\bf p}_a$ ($a= 1, \ldots , 8$)
at the vertices of a cube in momentum space,\cite{VolovikGorkov1985}
giving rise to four right-handed and four left-handed Weyl fermions.
In terms of the Cartesian unit vectors
($\widehat{\bf x}$, $\widehat{\bf y}$, $\widehat{\bf z}$),
the four Fermi points with right-handed Weyl fermions
($C_a=+1$, for $a= 1, \ldots , 4$) are given by
\begin{eqnarray}
{\bf p}_1 &=&{p_F\over \sqrt{3}}\;
(\widehat{\bf x}+\widehat{\bf y}+\widehat{\bf z})~,
\quad
{\bf p}_2  = {p_F\over \sqrt{3}}\;
(-\widehat{\bf x}-\widehat{\bf y}+\widehat{\bf z})~,
\nonumber\\
{\bf p}_3 &=&{p_F\over \sqrt{3}}\;
(\widehat{\bf x}-\widehat{\bf y}-\widehat{\bf z})~,
\quad
{\bf p}_4  = {p_F\over \sqrt{3}}\;
(-\widehat{\bf x}+\widehat{\bf y}-\widehat{\bf z})~,
\label{FourFermiPoints}
\end{eqnarray}
while the four Fermi points with
left-handed Weyl fermions ($C_a=-1$, for $a= 5, \ldots , 8$) have
opposite vectors. The vector ${\bf k}$ from all Fermi points vanishes,
since $q_a^2=1$ for the fermion charges $q_a=\pm 1$
and ${\bf p}_1+{\bf p}_2+{\bf p}_3+{\bf p}_4=0$ for the tetrahedron
(\ref{FourFermiPoints}).
The regular part of ${\bf k}$ also vanishes because of the discrete
cubic symmetry of the superconducting vacuum.

To the best of our knowledge, the $\alpha$--phase in superconductors
has not yet been established experimentally.
The example is, however, useful in that it
demonstrates that the total topological charge of the Fermi points
is zero for condensed-matter physics, $\sum_a N_a=0$.
For relativistic quantum field theory emerging near
Fermi points, this implies an equal number of right- and
left-handed fermions (some of which may have vanishing charges, though).

Consider, then, the $SU(3)_{\rm color}\times SU(2)_L\times U(1)_Y$
Standard Model\cite{EWSM} of elementary particle physics
above the electroweak transition (energy scale
$E_{\rm weak} \equiv G_{\rm Fermi}^{-1/2} \,(\hbar c)^{3/2}
\approx 300\,{\rm GeV}$).
For one family and with a right-handed neutrino included,
the model contains eight Fermi points with $N=+1$ and
eight Fermi points with $N=-1$, all located at ${\bf p}=0$.
The CPT-violating shifts ${\bf b}_a$ of the Fermi points may, in principle,
be different for the different species of fermions
($a=1,\ldots,16$), but gauge invariance
imposes certain restrictions; cf. Section~5 of Ref.~\refcite{CPTanomaly}.
Recall the representations of the sixteen left- and right-handed
fermions:
\begin{eqnarray}
L\; &:& \Bigl[\; (3,2)_{1/3} \;\Bigr]_{\rm quarks} +
            \Bigl[\; (1,2)_{-1}  \;\Bigr]_{\rm leptons}
\;,\nonumber\\[2mm]
R\; &:& \Bigl[\; (3,1)_{4/3}+ (3,1)_{-2/3} \;\Bigr]_{\rm quarks} +
            \Bigl[\; (1,1)_{-2}+ (1,1)_{0}     \;\Bigr]_{\rm leptons}       \;,
\label{SMirreps}
\end{eqnarray}
where the entries in parentheses denote $SU(3)$ and $SU(2)$
representations and the suffix the value of the hypercharge $Y$.
The electric charge $Q$ is given by the combination
$Y/2 + I_3$, with $I_3$ the weak
isospin from the diagonal Hermitian generator $T_3$ of the $SU(2)$
Lie algebra. The specific values of the hypercharge and  weak isospin are
$Y_a    \in \{ -2, -1, -2/3, 0, +1/3, +4/3 \}$ and
$I_{3a} \in \{ -1/2,0,+1/2 \}$.

For the $U(1)$ hypercharge gauge field of the Standard Model,
the charge $q_a$
in the sum (\ref{ManyFermiPointsSpaceLike2}) must be replaced by $Y_a$.
It could then be that the values of ${\bf p}_a$ have a
special pattern, similar to the pattern (\ref{FourFermiPoints})
of $\alpha$--phase superconductors, so that the
sum (\ref{ManyFermiPointsSpaceLike2}) vanishes exactly.
A complete analysis of this problem lies outside the scope of the
present article, but one simple solution can already be discussed.

Given the hypercharges $Y_a$ of the Standard Model (\ref{SMirreps}),
the following pattern of Fermi points may be assumed:
\begin{eqnarray}
{\bf p}^{(f)}_a &=& Y_a \;\,{\bf p}^{(f)}  \;,
\label{FermiPointsSM}
\end{eqnarray}
where, for the moment, the family index $f$ is set equal to 1.
The sixteen Fermi points from Eq.~(\ref{FermiPointsSM}) for $f=1$
depend  on a single vector, ${\bf p}^{(1)}$.
The factorized pattern (\ref{FermiPointsSM}) makes for a vanishing sum in
Eq.~(\ref{ManyFermiPointsSpaceLike2}) with $q_a$ replaced by
the hypercharge $Y_a$ and
also for the sum with $q_a$ replaced by the weak isospin $I_{3a}$.
Indeed, the very same sums occur for the perturbative gauge anomalies,
which are known to cancel out (cf. the last two papers of Ref.~\refcite{EWSM}).

With three known fermion families, $f=1,2,3$, the
pattern (\ref{FermiPointsSM}) is still a possibility. But, in general,
the basic vectors  ${\bf p}^{(f)}$ have different lengths,
\begin{equation}
\left|\,{\bf p}^{(f)} \right| \,\ne  \,
\left|\,{\bf p}^{(f')}\right|\;,
\quad {\rm for} \quad  f\neq f' \in \{ 1,2,3 \}\;,
\label{FermiPointsSMscales}
\end{equation}
and may or may not be aligned.
The pattern (\ref{FermiPointsSM})--(\ref{FermiPointsSMscales}), or
some other condition on the ${\bf p}^{(f)}_a$ with the same effect,
may indeed be required by the very tight experimental
limit\cite{CarollFieldJackiw}
on an electromagnetic \CS-like term,
$|\,{\bf k^{\rm exp}}| \lesssim 10^{-33}\,{\rm eV}/c$.
Without cancellations, one would need to explain the extreme
smallness of each value $|{\bf p}^{(f)}_a|$ separately.

For neutrinos, it is, in principle, possible that the splitting
mechanism is stronger than the mechanism leading to mass formation
(cf. Section~12.3.4 of Ref.~\refcite{VolovikBook}) and that
splitting occurs instead of mass generation.
Or one has both, but  with $|{\bf b}| > M$, so that the Fermi points
survive (cf. Section~\ref{sec:quantumphasetransition}).
The typical energy scale associated with the ${\bf p}^{(f)}_a$
of the massless (or nearly massless) neutrinos is perhaps
given by $(E_{\rm weak}/ E_{\rm Planck})^2 \times E_{\rm Planck}
\approx 10^{-5}\,{\rm eV}$,
if the corresponding CPT violation is a ``reentrance effect''
from nonsymmetric physics at the fundamental scale
$E_{\rm Planck} \equiv \sqrt{\hbar\,c^5/G}\,$, as discussed in
Section~12.4.3 of Ref.~\refcite{VolovikBook}.

The CPT-violating splittings
(\ref{FermiPointsSM})--(\ref{FermiPointsSMscales})
for the neutrinos have no direct consequences for the
electromagnetic sector, since the neutrinos are electrically neutral.
However, different splittings for the different species of neutrinos
may cause neutrino oscillations.
The basic idea is that there are three distinct
propagation states for the left-handed neutrinos
[cf. Eq.~(\ref{ModifiedEnergyDirac}) for $M=0$, with ${\bf b}$
replaced by the relevant $c\,{\bf p}^{(f)}_a$ and appropriate sign],
which need not be the same as the three interaction states.
For nontrivial mixing angles and large enough energy
of the initial neutrino (with definite momentum and flavor $f$),
the oscillation probabilities $P_{ff'}$ will be anisotropic but
essentially energy independent;
cf. Section~III B of Ref.~\refcite{ColemanGlashow}.
Note that this Fermi-point-splitting mechanism for neutrino oscillations
differs fundamentally from recent models based on CPT-violating neutrino
masses (see, e.g., Ref.~\refcite{BarenboimLykken} and references therein).

\section{Timelike Parameters and Fermi Surfaces}
\label{sec:Timelikeparameter}

The  Hamiltonian for a massive Dirac particle with an additional real
CPT-violating timelike parameter $b_0$ is
\begin{equation}
H=\left(\matrix{
      \vec{\sigma}\cdot ( c\,{\bf p}-{\bf b}) - b_0   &M\cr
M&- \vec{\sigma}\cdot ( c\,{\bf p}+{\bf b})  +b_0     \cr } \right) ~.
\label{ModifiedDiracTimeLike}
\end{equation}
The discrete symmetries of this Hamiltonian  have been studied,
for example, in Ref.~\refcite{Lehnert}.
For zero ${\bf b}$, the energy eigenvalues are given by
\begin{equation}
E^2_\pm= M^2+ \left(c\,|{\bf p}|\pm b_0 \right)^2  ~.
\label{ModifiedETimeLike}
\end{equation}
Topologically, this energy spectrum is similar to the one of
\BN~fermions in $s$--wave superconductors
where the role of the energy $M$ is played by the gap $\Delta$.
The spectrum (\ref{ModifiedETimeLike}) is fully gapped for $M\neq 0$,
but has a spherical Fermi surface of radius $|b_0|/c$
for $M=0$. This Fermi surface $|{\bf p}|=|b_0|/c$ is marginal,
as it consists of two identical Fermi surfaces with opposite global
topological invariants from the left- and right-handed fermions.
[The global topological invariant $N_a$ for the Fermi surface
$S_a$ is defined by Eq.~(\ref {TopInvariant}),
with a three-dimensional surface $\Sigma_a$ around  $S_a$
at $p_{0}=0$.]  The total topological charge describing the Fermi
surface $|{\bf p}|=|b_0|/c$ is thus zero and the Fermi surface is
not protected by topology (indeed, it disappears for $M\neq 0$).

\begin{figure}
\centerline{\psfig{file=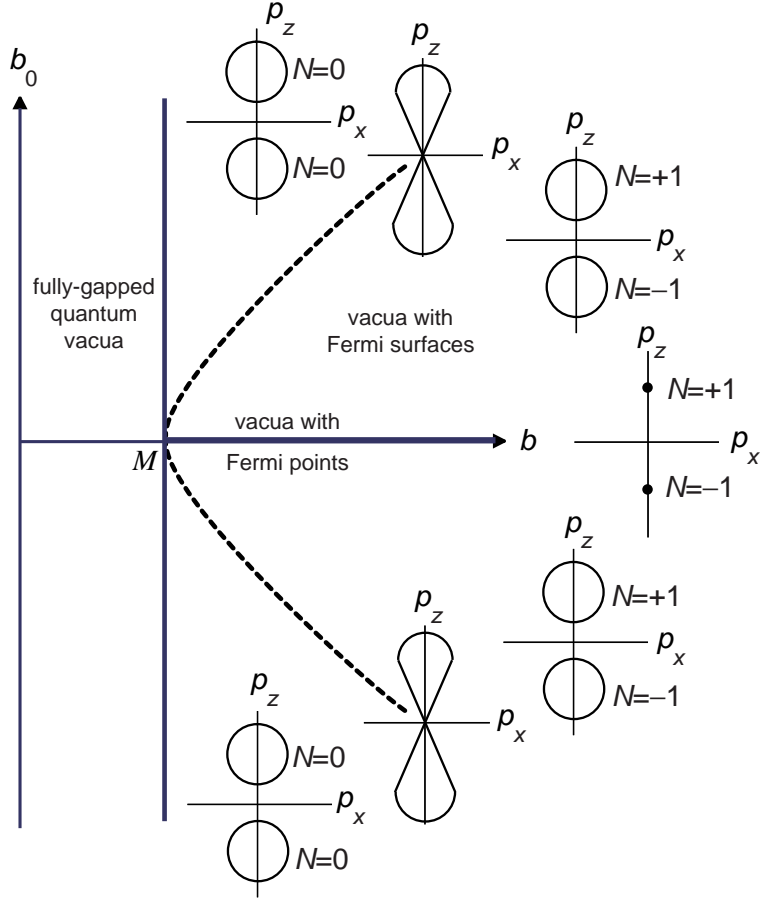,width=10cm}} 
\vspace*{8pt}
\caption{Phase diagram for parameters $b_0$ and $b \equiv |{\bf b}|$ of
the  modified Dirac Hamiltonian (\ref{ModifiedDiracTimeLike}).
For $b < M $, there are fully-gapped quantum vacua.
For $b \geq M$, there are vacua with Fermi surfaces
(shown schematically by the insets, with $p_y$ suppressed),
except for an open line segment $b > M$ at $b_0 =0$  with
topologically-protected Fermi points (shown by the  middle inset on the
right) and a line $b=M$ with topologically-unprotected Fermi points
(not shown by insets).
The values of the global topological invariants $N_a$ for the Fermi
surfaces $S_a$  or Fermi points ${\bf p}_a$ are indicated,
where $N_a$ is defined by Eq.~(\ref {TopInvariant}) with $\Sigma_a$
a three-dimensional surface around $S_a$ or ${\bf p}_a$ at
$p_{0}=0$. The dashed line $b^2-b_0^2=M^2$ (with Fermi surfaces meeting in
${\bf p} = {\bf 0}$) marks a quantum phase transition where the
global topological charges of the Fermi surfaces change
(see main text).}
\label{Fig2}
\end{figure}

Figure \ref{Fig2} clarifies the appearance of Fermi surfaces and
Fermi points in a $(b_0, |{\bf b}|)$ phase diagram.
The line $|{\bf b}|^2-b_0^2=M^2$ (with Fermi surfaces meeting in
${\bf p} = {\bf 0}$) marks a quantum phase transition where the
global topological charges $N_a$ of the Fermi surfaces change.
The Fermi surfaces of vacua with $|{\bf b}|^2-b_0^2>M^2$
inherit the topological charges $N=\pm 1$ from the Fermi points
of the line segment $|{\bf b}| > M$ at $b_0 =0$.
The Fermi surfaces of vacua with $0<|{\bf b}|^2-M^2<b_0^2$
have trivial global topology, $N=0$, and
shrink to points at $|{\bf b}|=M$.
These topologically-unprotected points with $N=0$ then
disappear for $|{\bf b}|<M$ (fully gapped vacua).
The  location of the quantum phase transition, $|{\bf b}|^2-b_0^2=M^2$,
is Lorentz-invariant. But the parameters $b_0$ and $|{\bf b}|$ are determined
in a preferred reference frame
(the frame of the heat bath in the context of condensed-matter physics),
which explains why other lines of the diagram change under boosts.

The analog of the Hamiltonian (\ref{ModifiedDiracTimeLike})
with massless ``relativistic'' fermions, $M=0$, occurs for
$^3$He--A; cf. Refs.~\refcite{VolovikBook} and \refcite{VolovikCPT}.
The flow of the superfluid component
with respect to the heat bath produces effective chemical potentials
$\mu_a$, which are opposite for left- and right-handed fermions.
After rescaling and deformation in momentum space, one has
\begin{equation}
H=\left(\matrix{
\vec{\sigma}\cdot ( c\,{\bf p}-{\bf b})- \bar{\mu} &0\cr
0&- \vec{\sigma}\cdot ( c\,{\bf p}+{\bf b})+ \bar{\mu}\cr}
\right)   ~,
\label{3HeATimeLike}
\end{equation}
where $\bar{\mu}\equiv \mu_1=-\mu_2=p_F\,{\bf v}_s\cdot \widehat{\bf l}$,
for a superfluid velocity ${\bf v}_s$ with respect to the heat bath.
The two Fermi points ${\bf p}_a=\pm {\bf b}/c$ for the line segment
$|{\bf b}|>0$
at $\bar{\mu}=0$ are transformed into two Fermi surfaces for $\bar{\mu}\neq 0$,
given by $| c\,{\bf p}- {\bf b}|=|\bar{\mu}|$ and
$| c\,{\bf p}+ {\bf b}|=|\bar{\mu}|$. For ${\bf b}\ne {\bf 0}$,
these two Fermi surfaces cannot cancel each other, as they do not  coincide.

According to  Eq.~(19.10) of Ref.~\refcite{VolovikBook},
the $k_0$ contribution to the induced \CSlike~term  (\ref{CSliketerm})
     has the Lagrange density
\begin{equation}
{\cal L}_{\rm CS-like}=- {1\over 8\pi^2}\;
{\bf A}\cdot ({\bf\nabla}\times {\bf A}) \sum_a \,C_a\,  q_a^2 \,\mu_a /c ~.
\label{ManyFermiPointsTimeLike}
\end{equation}
For $^3$He--A,
this term is responsible for the observed helical instability of the
``vacuum'' in the presence of
superflow.\cite{VolovikBook,VollhardtWoelfle} In general, it is known that a
\CSlike~term with timelike vector $k_\mu$, combined with the Maxwell
term, leads to vacuum instability;
cf. Refs.~\refcite{CarollFieldJackiw}
and \refcite{AdamKlinkhamerNPB}.  [Note that the
argument of Ref.~\refcite{AdamKlinkhamerPLB}  against a (timelike)
induced \CSlike~term from a fundamental CPT-violating \qfth~is simply
not applicable to the effective  theory of the
$^3$He--A liquid, since its ``vacuum''  can really be unstable.]

The Hamiltonian (\ref{3HeATimeLike}) does not mix
left- and right-handed fermions and is
equivalent to Eq.~(\ref{ModifiedDiracTimeLike}) for $M=0$,
if we are only interested
in the $k_0$ contribution to the \CSlike~term (\ref{CSliketerm}).
This implies that Eq.~(\ref{ManyFermiPointsTimeLike})
should also be valid for massless relativistic fermions (the
$k_0$ value determined by the Fermi surfaces does not depend on the
regularization; cf. Ref.~\refcite{VilenkinPRD}). Of course,
there could be an additional regular part of $k_{\mu}=(k_0,0,0,0)$,
which is determined by CPT violation at the fundamental level and which
does not depend on the presence of Fermi surfaces.

For the Standard Model fermions (\ref{SMirreps}),
there may be a pattern in the values of $b_{0a}^{(f)}$ of the
Hamiltonian (\ref{ModifiedDiracTimeLike}) for ${\bf b}=M=0$, which
nullifies the sum in Eq.~(\ref{ManyFermiPointsTimeLike}) with $\mu_a$
replaced by $b_{0a}^{(f)}$ and $q_a$ by $Y_a$ or $I_{3a}$.
As explained in the previous section, one example of such a pattern is
\begin{equation}
b_{0a}^{(f)} = Y_a \; b_{0}^{(f)}\;,\quad
\left|\,b_{0}^{(f)} \right| \,\ne  \,
\left|\,b_{0}^{(f')}\right|\;,
\quad {\rm for} \quad  f\neq f' \in \{ 1,2,3 \}\;.
\label{b0pattern}
\end{equation}
Different values of $b_{0a}^{(f)}$ for the different species of
neutrinos may again cause neutrino oscillations
[cf. Eq.~(\ref{ModifiedETimeLike}) for $M =0$, with $b_0$
replaced by the relevant $b_{0a}^{(f)}$ and appropriate sign].

\section{Finite-Size Effects and Defects}
\label{sec:finite-size}

In the context of \rqfth~(specifically, \cgth),
an anomalous \CSlike~term has been found to
originate from nontrivial space
topology.\cite{CPTanomaly,FundamentalTimeAsymmetry,RigorousResult,SpacetimeFoam}
For the present article, the splitting of Fermi points
or the appearance of a Fermi surface (both of which induce
a  \CS~vector ${\bf k}$) may also be caused by finite-size effects
at the fundamental level or by defects in the fabric of space.
If so, ${\bf b}$ would be either inversely proportional to
the size of the compact dimension,
$|{\bf b}| \propto 1/L$ (cf. Ref.~\refcite{CPTanomaly}), or
depend on the distance
from the space defect (cf. Ref.~\refcite{SpacetimeFoam}).

In condensed-matter physics, an interesting example
is given by a combined topological defect in $^3$He--A, namely
vortex--disgyration. Its vortex part is characterized by the
circulating superfluid velocity
${\bf v}_s({\bf r})=\hbar/(2m_3\,\rho)\:\widehat{\mbox{\boldmath $\phi$}}$,
with $\rho$ the distance from the
vortex axis and $m_3$  the mass of the $^3$He atom. The disgyration is
represented by the azimuthal arrangement of the $\widehat{\bf l}$
field, $\widehat{\bf l}({\bf r})= \widehat{\mbox{\boldmath $\phi$}}$.
This type of topological texture occurs in the asymptotic region of the
so-called circular Mermin--Ho vortex; cf. Fig. 1a (right) of
Ref.~\refcite{ZotosMaki}. The vortex produces the component
$b_0=p_F\,{\bf v}_s({\bf r})\cdot\widehat{\bf l}({\bf r})=\hbar\,
p_F/(2m_3\,\rho)$ and the azimuthal disgyration gives the vector
${\bf b}= \hbar \, p_F /(2m_3\,\rho)\:\widehat{\mbox{\boldmath $\phi$}}$.
This last vector induces a space-dependent \CS~vector ${\bf k}({\bf r})$ of the
type discussed in Section~II C of Ref.~\refcite{SpacetimeFoam}.

\section{Conclusion}
\label{sec:Conclusion}

Timelike and spacelike CPT-violating parameters $b_\mu$  in the fermionic
sector lead to different momentum-space topology and, hence, to
a different response of the parameters
$k_\mu$ of the \CSlike~term in the effective gauge-field action. All three
universality classes of fermionic vacua (with mass gaps, Fermi points,
or Fermi surfaces) can play a role. The induced ``vector'' $k_\mu$ depends
on the particular universality class chosen by the system.
These observations are also relevant to elementary particle physics,
both if the Standard Model is an emergent
phenomenon of a fermionic quantum vacuum or if the Standard Model
is part of a fundamental theory.
Of particular interest may be a new CPT-violating mechanism for neutrino
oscillations, as discussed in the last paragraphs of Sections
\ref{sec:p-wavesuperconductors} and \ref{sec:Timelikeparameter}.

A purely spacelike vector $k_\mu=(0,{\bf k})$ in the CPT-violating
\CSlike~action (\ref{CSliketerm}) consists of two parts.
The regular part, ${\bf k}^{\rm reg}$, can be nonzero due to explicit
CPT violation at the fundamental level. For the concrete case of
the  Dirac Hamiltonian (\ref{ModifiedHDiracGauge}),
${\bf k}^{\rm reg}$ is a regular function of the
CPT-violating parameter ${\bf b}$.
That is, ${\bf k}^{\rm reg}$ is continuous
across the quantum phase transition at $b \equiv |{\bf b}| = M$
between the vacua of the different
universality classes.   The anomalous (nonanalytic) part,
${\bf k}^{\rm anom}$, comes solely from the Fermi points (present
for $b > M$)  and is proportional to their splitting. This
nonanalytic contribution is regularization independent. For the case of a
single massive Dirac  fermion, the vector
${\bf k}^{\rm anom}$ is given by Eq.~(\ref{SpaceLikeKAphase}).

Moreover, the general regularization-independent expression
(\ref{ManyFermiPointsSpaceLike})
for ${\bf k}^{\rm anom}$ depends only on the topological properties
of the Fermi points (cf. \ref{sec:Appendix})
and is applicable to any system
which contains one or more pairs of Fermi points, even if the
system does not obey relativistic invariance.
The derivation of our expression used only gauge invariance.
This is one of many examples where
topological effects do not require the mathematics of relativistic
invariance. Indeed, the result is applicable to nonrelativistic $^3$He--A
as well.

The example of $^3$He--A demonstrates that there may be a regular part,
${\bf k}^{\rm reg}$, which does not depend on the existence of Fermi
points.  This part comes from the angular momentum of the liquid
(corresponding to spontaneously broken time-reversal symmetry)
and is the same with or without Fermi points.  In contrast, the nonanalytic
part, ${\bf k}^{\rm anom}$, comes from anomalies generated by the Fermi
points. The total spacelike \CSlike~term, with vector
${\bf k}\propto p_F\,{\bf l}$, is responsible for the orbital dynamics
(i.e., the dynamics of the
unit vector $\widehat{\bf l}$ directed along the orbital angular
momentum of the Cooper pairs).

Real $^3$He--A lives deep inside the Fermi-point region of the phase
diagram where the regular and nonanalytic terms almost cancel each other.
This last circumstance impacts on many phenomena of
$^3$He--A including the dynamics of topological defects, e.g., quantized
vortices. There would be no Fermi points and related anomalies
on the other side of the quantum phase transition.

This region of anomaly-free vacua, as well as the quantum phase
transition between vacua with Fermi points and fully-gapped
vacua, can perhaps be probed with laser-manipulated Fermi gases.
Recently, the condensation of pairs of fermionic $^{40}$K atoms  has been
reported\cite{FermionicAtomPairs} in the crossover regime
(the so-called BEC--BCS crossover). In this experiment,
a magnetic-field Feshbach resonance was
used to control the interactions in the Cooper channel.
But the system considered has (spin-singlet) $s$--wave pairing
and there are fully-gapped vacua on both sides of the crossover,
without quantum phase transition.

For the case of pairing interactions in the
(spin-triplet) $p$--wave channel,
the  modulus $b$ of the parameter ${\bf b}$ in the
Hamiltonian (\ref{ModifiedHDirac}) can be taken to be proportional to the
inverse strength of the atom--atom interactions in the $p$--wave channel
and its direction $\widehat{{\bf b}} \equiv  \widehat{\bf z}$
to be along the unit vector $\widehat{\bf l}$ of the orbital angular
momentum of the Cooper pairs. We may, then, expect the BEC--BCS crossover
to be accompanied by the quantum phase transition of
Fig.~\ref{Fig1}, with a change in the momentum-space topology of the
vacuum and the universality class.
Similar quantum phase transitions may be relevant to Standard
Model physics.

\section*{Acknowledgements}

The work of G. E. Volovik is supported in part
by the Russian  Foundation for Fundamental Research
under grant $\#$02-02-16218
and by the Russian Ministry of Education and
Science, through the Leading Scientific School grant $\#$2338.2003.2 and
the Research Program  ``Cosmion.'' This work is also supported by the
European Science Foundation COSLAB Program.

\appendix
\section{Momentum-Space Topological Invariant}
\label{sec:Appendix}

In this appendix, we discuss the origin of the gauge invariance of
the induced Chern--Simons-like term,
even though it contains explicitly the gauge potential ${\bf A}$. The
Chern--Simons-like term (\ref{CSliketerm}) becomes gauge invariant if,
for a special reason, the Chern--Simons vector
${\bf k}$ does not depend on space or time. There must then be a
protection mechanism which keeps the vector ${\bf k}$ constant
under perturbations of the system. Such a protection mechanism is well known in
condensed-matter systems, where
it has a topological origin. Certain physical parameters of
these systems, such as the Hall and the spin-Hall conductivity, are robust to
perturbations, because they can be expressed in terms of momentum-space
invariants or more complicated momentum-space invariants protected
by symmetry; cf. Refs.~\refcite{VolovikBook} and \refcite{Yakovenko2}.
The question, now, is if topological
protection also operates for the anomalous contributions to the
Chern--Simons  vector ${\bf k}$ coming the Fermi points.

We start by rewriting the momentum-space topological invariant
$N_a$, as defined by Eq.~(\ref{TopInvariant}) in terms of the fermionic
propagator $G(ip_0,{\bf p})$. For this purpose, we introduce a
matrix density in four-dimensional energy-momentum space,
\begin{equation}
{\cal N}_\nu \equiv
   {1\over{24\pi^2}}\;e_{\kappa\lambda\mu\nu}~
    G\frac{\partial}{\partial p_\kappa} G^{-1}\;
    G\frac{\partial}{\partial p_\lambda}  G^{-1}\;
    G\frac{\partial}{\partial p_\mu}G^{-1}~.
\label{TopInvariantDensity}
\end{equation}
With this matrix density,  we have the following expressions for the
topological charge $N_a$ of the particular Fermi point with label $a$
and for the general density of topological charge:
\begin{eqnarray}
N_{a} &=& {\rm tr}\, \oint_{\Sigma_a} dS^{\nu}\;{\cal N}_\nu~,
\nonumber\\
{\rm tr}\;\frac{\partial}{\partial p_\nu}{\cal N}_\nu &=&
\sum_a N_a \,\prod_{\mu=0}^{3}\delta\left(p_\mu - p_{\mu a}\right)~.
\label{TopInvariant2}
\end{eqnarray}
Here, $\Sigma_a$ is a three-dimensional surface around the isolated
Fermi point $p_{\mu a}= (\,0,{\bf p}^{(0)}_a\,)$
and `tr' stands for the trace over all indices of the Green's function,
normalized by ${\rm tr}\, \identity = 2\,N_{\rm Weyl}$,
with $N_{\rm Weyl}$ the number of two-component Weyl spinors.

Equations (\ref{TopInvariantDensity}) and (\ref{TopInvariant2})
make clear
that the anomalous part of the Chern--Simons vector $k_\mu=(0,{\bf k})$ in
Eq.~(\ref{ManyFermiPointsSpaceLike}) can be written as  follows:
\begin{equation}
k_\mu= {1\over{24\pi^2}}\;{\rm tr}\; \int d^4 p\;\; p_\mu  \;
       {\cal Q}^2\; \frac{\partial}{\partial p_\nu}{\cal N}_\nu  ~,
\label{GeneralSpaceLike}
\end{equation}
where  the integral is over the whole four-dimensional energy-momentum
space and  ${\cal Q}$ is the matrix of the charges ($q_a$)
of the fermions interacting with the $U(1)$ gauge field
${\bf A}$. This matrix of charges commutes with the
Green's function matrix provided the $U(1)$ gauge symmetry holds.
Integrating by parts and assuming the matrices ${\cal N}_\nu$ to
vanish fast enough towards infinity, one obtains
\begin{equation}
k_\mu= - {1\over{24\pi^2}}\;{\rm tr}\; \int d^4 p\;
    \;{\cal Q}^2 \;{\cal N}_\mu  ~.
\label{GeneralSpaceLike2}
\end{equation}
A surface integral must be added, if the matrices ${\cal N}_\nu$ do
not vanish fast enough.

It can be shown that Eq.~(\ref{GeneralSpaceLike2}) is
invariant under perturbations of the Green's function $G$ which do not
violate the $U(1)$ gauge symmetry and do not
change the positions of the singular points.
The restricted topological stability of the Fermi-point contribution
to the vector ${\bf k}$ shows
that this contribution cannot depend on space and time.
The corresponding Chern--Simons-like term (\ref{CSliketerm}) is
thus gauge invariant.

The robustness of the anomalous vector ${\bf k}$ under
variations of the Green's function implies that it
does not depend on changes of the ultraviolet cutoff.
Naively, we do not expect a dependence on the infrared cutoff either.
Hence, the anomalous
part of ${\bf k}$ does not undergo radiative corrections.  This
suggests a type of
``nonrenormalization theorem'' for ${\bf k}$, which resembles the
Adler--Bardeen theorem for the chiral anomaly.\cite{AdlerBardeen,Adler}

Similar integrals which are invariant under restricted deformations of the
Green's function give rise to quantization of physical parameters in
(2+1)-dimensional condensed-matter systems, examples being the Hall and
spin-Hall conductivity.
These parameters are expressed in terms of momentum space integrals
which are invariant under deformations preserving a particular symmetry of
the system  (see, e.g., Sections~12.3.1 and 21.2.3 in Ref.~\refcite{VolovikBook}
and Ref.~\refcite{Yakovenko2}).
In all these cases, the Chern--Simons terms can be written as
the product of coordinate-space and momentum-space topological terms.
Essentially the same holds for Eqs.~(\ref{CSliketerm})
and (\ref{GeneralSpaceLike2}) in $3+1$ dimensions.

For (3+1)-dimensional condensed-matter systems,
the Chern--Simons-like term (\ref{CSliketerm}) with
vector (\ref{GeneralSpaceLike2}) suggests that there may be a
nonanalytic contribution to the Hall conductivity tensor
$\sigma_{\rm H}^{mn}$ in bulk nonsuperconducting materials.
With the definition $j^m=\sigma_{\rm H}^{mn}\,E_n$,
for the electric field ${\bf E}$, we find:
\begin{equation}
{\bf j}={\delta S_{\rm \,CS-like}\over \delta
{\bf A}}= 2\,{\bf k} \times {\bf E}\;, \quad
\sigma_{\rm H}^{mn}=2\,\epsilon^{mnl}\:k_l  ~.
\label{3dHall}
\end{equation}
This contribution from the Fermi points comes in addition to
the regular Hall conductivity of three-dimensional systems as discussed in
the literature  (see, e.g., Ref.~\refcite{Hasegawa} and references therein).
The regular vector ${\bf k}^{\rm reg}$ in  fully-gapped systems
is determined by one of the vectors ${\bf G}$ on
the reciprocal lattice in crystals or in spin/charge density waves,
${\bf k}^{\rm reg}=(e^2/8\pi^2 \hbar)\,{\bf G}$; cf. Ref.~\refcite{Halperin}.
For the anomalous vector ${\bf k}$ in these periodic systems,
the spatial part of the four-dimensional volume
in Eq.~(\ref{GeneralSpaceLike2}) includes
the elementary cell of the reciprocal lattice (Brillouin zone)
and `tr' has a trace over the band indices.


\end{document}